\documentclass[a4paper]{article}
\usepackage{wqit}
\usepackage{mathrsfs,amsmath,amssymb}\input{Qcircuit}
\def\sp#1{\mathscr #1}
\def\N#1{|\!|#1|\!|}

\def\transp#1{{#1}^\tau}
\def\SU#1{\mathbb{SU}(#1)}
\def\sH{\sp H}
\def\set#1{\mathscr #1}

\def\Reals{\mathbb R}
\def\Cmplx{\mathbb C}
\def\<{\langle}\def\>{\rangle}
\def\bP{{\bf P}}\def\bE{{\bf E}}\def\bQ{{\bf Q}}\def\bF{{\bf F}}
\def\Pn{\mathscr{P}_n}\def\cP{\mathscr{P}}
\def\anc{\mathscr{A}}
\def\map#1{\mathscr{#1}}\def\dim{\operatorname{dim}}

\def\Unt#1{\set{U(#1)}}
\def\transp#1{{#1}^\intercal}
\def\Tr{\operatorname{Tr}}

\def\N#1{\left|\!\left|#1\right|\!\right|}
\def\eg{e.~g. }
\def\conv#1{{\mathcal #1}}

\begin{document}

\title{Programmable quantum channels and measurements}

\author{
  Giacomo Mauro D'Ariano 
  \affiliation{1}
  \email{dariano@unipv.it}
  \and 
   Paolo Perinotti
  \affiliation{1}
  \affiliation{2}
  \email{perinotti@fisicavolta.unipv.it}
}
\address{1}{QUIT group, Dipartimento di Fisica ``A. Volta'',
via Bassi 6, I-27100, Pavia, Italy}
\address{2}{CNR-INFM, Unit\`a di Pavia}

\abstract{We review some partial results for two strictly related
  problems. The first problem consists in finding the optimal joint
  unitary transformation on system and ancilla which is the most
  efficient in programming any desired channel on the system by
  changing the state of the ancilla. In this respect we present a
  solution for $\dim(\sH)=2$ for both system and ancilla. The second
  problem consists in finding the optimal universal programmable
  detector, namely a device that can be tuned to perform any desired
  measurement on a given quantum system, by changing the state of an
  ancilla. With a finite dimension $d$ for the ancilla only
  approximate universal programmability is possible, with $d=d(\varepsilon^{-1})$
  increasing function versus $\varepsilon^{-1}$. We show 
  that one can achieve $d(\varepsilon^{-1})$ polynomial, and even linear in specific cases.}


\maketitle

\section{Introduction}\footnote{Work Presented at Workshop on Quantum Information Theory and Quantum Statistical Inference, 17-18 November 2005, Tokyo, ERATO Quantum Computation and Information Project}
A fundamental problem in quantum computing and, more generally, in
quantum information processing \cite{Nielsen} is to experimentally
achieve any theoretically designed quantum channel or detector using a
fixed device, through a suitable program encoded in the state of an
ancillary system. While a large branch of theoretical research in
quantum information addressed the design of algorithms and of circuits
to solve precise problems, a parallel research line is that of designing devices that can be
programmed to achieve different tasks, just like classical computers do. Moreover, designing a programmable
quantum gate or detector is a problem of relevance for example in
proving the equivalence of cryptographic protocols, \eg proving the
equivalence between a multi-round and a single-round quantum bit
commitment \cite{dsw}, or when trying to eavesdrop quantum-encrypted
information. What makes the problem of gate programmability non
trivial is that exact universal programmability of channels is
impossible, as a consequence of a no-go theorem for programmability of
unitary transformations by Nielsen and Chuang \cite{niels}. A similar
situation occurs for universal programmability of
POVM's \cite{fiurasek2,our}. In both cases, it is still possible to
achieve programmability probabilistically \cite{niels,buzekproc,buzek},
or even deterministically \cite{ciravid}, though within some accuracy.
In establishing the theoretical limits to state-programmability of
channels or POVM's the starting problem is to find the joint
system-ancilla unitary or observable, respectively, which achieves the
best accuracy for fixed dimension of the ancilla: this is exactly the
problem that is addressed in the present paper. This problem turned out
to be hard, even for low dimension. Here we will give a solution for
the optimal device for programming unitary channels for dimension two 
for both system and ancilla. On the other hand, as regards programming observables,
we will give an upper bound for the optimal ancilla dimension $d(\varepsilon^{-1})$ 
versus the accuracy $\varepsilon{-1}$ for programmable detectors. As we will see, it turns
out \cite{our} that a dimension $d(\varepsilon^{-1})$ increasing polynomially with precision
$\varepsilon^{-1}$ is possible, and even a linear dependence is achievable for specific cases. 
This should be compared with the preliminary indications of an exponential growth of
Ref. \cite{fiurasek1}. However, even the linear dependence $d(\varepsilon^{-1})$ is still suboptimal.

\section{Statement of the problems}

\paragraph{Programmable unitaries}

We want to program unitary channels by a fixed device as follows
\begin{equation}\label{partrace}
\map{P}_{V,\sigma}(\rho)\doteq\Tr_2[V(\rho\otimes\sigma)V^\dag],
\end{equation}
with the system in the state $\rho$ interacting with an ancilla in the
state $\sigma$ via the unitary operator $V$ of the programmable device
(the state of the ancilla is the {\em program}).  For fixed $V$ the
above map can be regarded as a linear map from the convex set of the
ancilla states $\conv{A}$ to the convex set of channels for the system
$\conv{C}$. We will denote by $\conv{P}_{V,\conv{A}}$ the image of the
ancilla states $\conv{A}$ under such linear map: these are the
programmable channels. According to the well known no-go theorem by
Nielsen and Chuang it is impossible to program all unitary channels on
the system with a single $V$ and a finite-dimensional ancilla, namely
the image convex $\conv{P}_{V,\conv{A}}\subset\conv{C}$ is a proper
subset of the whole convex $\conv{U}$ of unitary channels and their
convex combinations. This opens the following problem:
\begin{itemize}
\item[] {\bf Problem:} {\em For given dimension of the ancilla, find
    the unitary operators $V$ that are the most efficient in
    programming unitary channels, namely which minimize the largest
    distance $\varepsilon(V)$ of each channel $\map{U}\in{\conv{U}}$
    from the programmable set $\conv{P}_{V,\conv{A}}$: }
\begin{equation}\label{eps}
\varepsilon(V)\doteq\max_{\map{U}\in{\conv{U}}}\min_{\map{P}\in\conv{P}_{V,\conv{A}}}
\delta(\map{C},\map{P})\equiv\max_{\map{U}\in{\conv{U}}}\min_{\sigma\in\conv{A}}
\delta(\map{C},\map{P}_{V,\sigma}).
\end{equation}
\end{itemize}
\medskip\par As a definition of distance it would be most appropriate
to use the CB-norm distance $\N{\map{C}-\map{P}}_{CB}$. However, this
leads to a very hard problem. We will use instead the following
distance
\begin{equation}\label{del}
\delta(\map{C},\map{P})\doteq \sqrt{1-F(\map{C},\map{P})},
\end{equation}
where $F(\map{C},\map{P})$ denotes the Raginsky fidelity
 \cite{raginski}, which for unitary map $\map{C}\equiv\map{U}=U\cdot
U^\dag$ is equivalent to the channel fidelity  \cite{Nielsen}
\begin{equation}\label{ragfidel}
F(\map{U},\map{P})=\frac{1}{d^2}\sum_i|\Tr[C_i^\dag U]|^2,
\end{equation}
where $\map{C}=\sum_i C_i\cdot C_i^\dag$. Such fidelity is also
related to the input-output fidelity averaged over all pure states
$\overline{F}_{io}(\map{U},\map{P})$, by the formula
$\overline{F}_{io}(\map{U},\map{P})=[1+dF(\map{U},\map{P})]/(d+1)$.
Therefore, our optimal unitary $V$ will maximize the fidelity
\begin{equation}\label{ragfidel2}
F(V)\doteq\min_{U\in\Unt{H}}F(U,V),\quad F(U,V)\doteq\max_{\sigma\in\conv{A}}
F(\map{U},\map{P}_{V,\sigma}) 
\end{equation}

\paragraph{Programmable detectors}

The POVM of a measuring apparatus is a set of positive operators $P_i\geqslant0$, $i=1,\ldots n$,
$n<\infty$ normalized to the identity $\sum_{i=1}^nP_i=I$, which gives the probability distribution
of the outcomes for each input state $\rho$ via the Born rule
\begin{equation}
p(i|\rho)\doteq\Tr[\rho P_i].
\end{equation}
Clearly, the most general programmable detector is described by an
observable $\bF\doteq\{F_i\}$ jointly measured on system and ancilla. The probability distribution
of the outcomes is given by
\begin{equation}
p_\sigma(i|\rho)=\Tr[(\rho\otimes\sigma) F_i],\;\forall i,\forall\rho.
\label{eq:mathdescr}
\end{equation}
By taking the partial trace in Eq. \eqref{eq:mathdescr} over the ancilla and using the polarization
identity (Eq.  \eqref{eq:mathdescr} holds for all states) one obtains the POVM
\begin{equation}
P_{\sigma,i}=\Tr_2[(I\otimes\sigma)F_i].
\label{eq:prog}
\end{equation}
From Eq. (\ref{eq:prog}) it follows that the convex set of
states $\anc$ of the ancilla is in correspondence via the map
$\bP_{\bF,\sigma}\doteq\Tr_2[(I\otimes\sigma)\bF]$ with a convex
subset $\cP_{\bF,\anc}\subseteq\Pn$ of the convex set $\Pn$ of the system POVM's
with $n$ outcomes. The symbol $\cP_{\bF,\anc}$ denotes the convex set of programmable 
POVM's that can be achieved with fixed observable $\bF$ and varying state $\sigma\in\anc$.
The no-go theorem proved in  \cite{fiurasek1,our} states that for any fixed observable $\bF$ the
programmable set $\bP_{\bF,\anc}$ is strictly contained in $\Pn$, since even just the observables
cannot be programmed with a fixed observable $\bF$ and a finite dimensional ancilla. 
We now restrict attention to programmability of observables only, whence $n\equiv\dim(\sH)$, the
case of nonorthogonal POVM's simply resorting to program observables on a larger Hilbert space.
In the following we will denote by ${\cal O}_n$ the set of observables. 
 The problem of measurement programmability
can then be stated in mathematical terms as follows
\begin{itemize}
\item[] {\bf Problem:} {\em For given dimension of the ancilla, find
    the joint observables $\bF$ that are the most efficient in
    programming system observables, namely which minimize the largest
    distance $\varepsilon(\bF)$ of each observable $\bP\in{\cal O}_n$
    from the programmable set $\bP_{\bF,\sigma}$: }
\begin{equation}\label{minmax}
\varepsilon(\bF)\doteq\max_{\bP\in{\cal O}_n}\min_{\bQ\in\bP_{\bF,\sigma}}
\delta(\bP,\bQ)\equiv\max_{\bP\in{\cal O}_n}\min_{\sigma\in\conv{A}}
\delta(\bP,\bP_{\bF,\sigma}).
\end{equation}
\end{itemize}
We define the distance between two POVM's as the distance between
their respective probabilities, maximized over all possible states,
namely
\begin{equation}
\delta({\bP},{\bQ})=\max_\rho\sum_i|\Tr[\rho(P_i-Q_i)]|\,.
\label{eq:dist}
\end{equation}
The distance defined in Eq. \eqref{eq:dist} is hard to handle
analytically, whence we bound it as follows
\begin{equation}
\delta(\bP,\bQ)\leqslant\sum_i\N{P_i-Q_i}\leqslant\sum_i\N{P_i-Q_i}_2,
\label{eq:normbound}
\end{equation}
where $\N{A}$ is the usual operator norm of $A$, and
$\N{A}_2\doteq\sqrt{\Tr[A^\dag A] }$ is the Frobenius norm.
\medskip\par

\section{Programming qubit unitaries}

By some lengthy calculation we can obtain the Kraus operators for the
map $\map{P}_{V,\sigma}(\rho)$
\begin{equation}
\begin{split}
  &\map{P}_{V,\sigma}(\rho)=\sum_{nm}C_{nm}\rho C_{nm}^\dag,\\
  &C_{nm}=\sum_ke^{i\theta_k}\Psi_k|\upsilon_n^*\>\<\upsilon_m^*|\Psi_k^\dag\sqrt{\lambda_m}
\end{split}
\end{equation}
where $|\upsilon_n\>$ denotes the eigenvector of $\sigma$
corresponding to the eigenvalue $\lambda_n$ and $^*$ denotes complex
conjugation in the same fixed basis for which the operator $\Psi_k$ have the same matrix elements
as the matrix of coefficients of the eigenvector of $V$ corresponding to eigenvalue $e^{i\theta_k}$.
We then obtain
\begin{equation}
\begin{split}
\sum_{nm}|\Tr[C_{nm}^\dag U]|^2=&
\sum_{kh}e^{i(\theta_k-\theta_h)}\Tr[\Psi_k^\dag U^\dag\Psi_k\transp{\sigma}\Psi_h^\dag U\Psi_h]\\
=&\Tr[\transp{\sigma} S(U,V)^\dag  S(U,V)]
\end{split}
\end{equation}
where 
\begin{equation}
S(U,V)=\sum_k e^{-i\theta_k}\Psi^\dag_kU\Psi_k\,.\label{esse}
\end{equation}
and $\transp{}$ denotes transposition in the canonical basis. The
fidelity (\ref{ragfidel2}) can then be rewritten as follows
\begin{equation}\label{avfuv}
F(U,V)=\frac1{d^2}\N{S(U,V)}^2.
\end{equation}
The operator $S(U,V)$ in Eq. (\ref{esse}) can be written as follows
\begin{equation}
S(U,V)=\Tr_1[(\transp{U}\otimes I)V^*]\,.
\end{equation}
Changing $V$ by local unitary operators transforms $S(U,V)$ in the
following fashion
\begin{equation}
S(U,(W_1\otimes W_2)V(W_3\otimes W_4))=W_2^*S(W_1^\dag UW_3^\dag,V)W_4^*,
\end{equation}
namely the local unitaries do not change the minimum fidelity, since
the unitaries on the ancilla just imply a different program state,
whereas the unitaries on the system just imply that the minimum
fidelity is achieved for a different unitary---say $W_1^\dag
UW_3^\dag$ instead of $U$.\par

\begin{figure}[h]
\begin{center}
  $$\Qcircuit @C=1em @R=.7em {  & \ctrl{1} & \gate{X_{\pm\frac{\pi}{4}}} & \ctrl{1} & \qw \\
    & \targ & \gate{Z_{\mp\frac{\pi}{4}}} & \targ & \qw }$$
\caption{Quantum circuit scheme for the optimal unitary operator $V$ in Eq. (\ref{optV}). $W_\alpha$ denotes $e^{i\frac\alpha2\sigma_W}$. For the derivation of the circuit consider that $\sigma_x\otimes\sigma_x=C (\sigma_x\otimes I)C$ and $\sigma_z\otimes\sigma_z=C (I\otimes\sigma_z)C$, where $C$ denotes the controlled-not.
\label{circuitV}}
\end{center}
\end{figure}
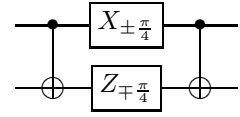

For system and ancilla both two-dimensional, one can parameterize all
possible joint unitary operators as follows \cite{KrausCirac,glaser}
$(W_1\otimes W_2)\tilde V (W_3\otimes W_4)$, where
\begin{equation}
\tilde V=\exp[i(\alpha_1\sigma_1\otimes\transp{\sigma_1}+\alpha_2\sigma_2\otimes\transp{\sigma_2}+\alpha_3\sigma_3\otimes\transp{\sigma_3})]\,.
\label{parambipu}
\end{equation}
The problem is now reduced to study only joint unitary operators of
the form of Eq.~\eqref{parambipu}. It can be proved that the
coefficients of its eigenvectors are the matrix elements of Pauli
matrices $\sigma_j$, $j=0,1,2,3$ where $\sigma_0=I$,
$\sigma_1=\sigma_x$, $\sigma_2=\sigma_y$, $\sigma_3=\sigma_z$. This
means that we can rewrite $S(U,V)$ in Eq.~\eqref{esse} as follows
\begin{equation}\label{suvqub}
S(U,V)=\frac12\sum_{j=0}^3e^{-i\theta_j}\sigma_j U\sigma_j\,,
\end{equation}
with
\begin{equation}
\theta_0=\alpha_1+\alpha_2+\alpha_3\,,\quad\theta_i=2\alpha_i-\theta_0\,.
\end{equation}
Through the derivation described in Appendix \ref{appbloch} we obtain
that the fidelity minimized over all unitaries is given by
\begin{equation}
F(V)=\frac1{d^2}\min_j|t_j|^2.
\end{equation}
where 
\begin{equation}
\begin{split}
  &t_0=\frac12\sum_{j=0}^3e^{-i\theta_j},\\
  &t_j=e^{-i\theta_0}+e^{-i\theta_j}-t_0,\;1\leq j\leq 3.
\end{split}
\end{equation}
The optimal unitary $V$ is now obtained by maximizing $F(V)$. We need
then to consider the decomposition Eq.~\eqref{parambipu}, and then to
maximize the minimum among the four eigenvalues of $S(U,V)^\dag
S(U,V)$.  Notice that $t_j=\sum_{\mu}H_{j\mu}e^{i\theta_\mu}$, where
$H$ is the Hadamard matrix
\begin{equation}
H=\frac12
\begin{pmatrix}
1&1&1&1\\
1&1&-1&-1\\
1&-1&1&-1\\
1&-1&-1&1
\end{pmatrix},
\end{equation}
which is unitary, and consequently $\sum_j|t_j|^2=\sum_j
|e^{i\theta_j}|^2=4$.  This implies that $\min_j|t_j|\leq1$. We now
provide a choice of phases $\theta_j$ such that $|t_j|=1$ for all $j$,
achieving the maximum fidelity allowed. For instance, we can take
$\theta_0=0,\theta_1=\pi/2,\theta_2=\pi,\theta_3=\pi/2$, corresponding
to the eigenvalues $1,i,-1,i$ for $V$. Another solution is
$\theta_0=0,\theta_1=-\pi/2,\theta_2=\pi,\theta_3=-\pi/2$. Also one
can set $\theta_i\to -\theta_i$. The eigenvalues of $S(U,V)^\dag
S(U,V)$ are then $1,1,1,1$, while for the fidelity we have
\begin{equation}\label{optV}
F\doteq\max_{V\in\Unt{H^{\otimes 2}}}F(V)=\frac{1}{d^2}=\frac{1}{4},
\end{equation}
and the corresponding optimal $V$ has the form
\begin{equation}
V=\exp\left[\pm i\frac{\pi}{4}\left(\sigma_x\otimes\sigma_x\pm\sigma_z\otimes\sigma_z\right)\right].
\end{equation}
A possible circuit scheme for the optimal $V$ is given in Fig.
\ref{circuitV}.

\par 
Such fidelity cannot be achieved by any $V$ of the controlled-unitary form
\begin{equation}
V=\sum_{k=1}^2V_k\otimes|\psi_k\>\<\psi_k|,\qquad \<\psi_1|\psi_2\>=0,
\end{equation}
where $V_1, V_2$ are unitaries on $\sH\simeq\Cmplx^2$. In fact, it is easy to see that in this case
the fidelity is given by
\begin{equation}
F(U,V)=\frac{1}{4}|\Tr[V_h^\dag U]|^2,\qquad h=\arg\max_k|\Tr[V_k^\dag U]|,
\end{equation}
and for any couple of unitaries $\{V_k\}$ there always exists a unitary $U$ orthogonal to both
$\{V_k\}$, whence $F(V)\doteq\min_{U\in\Unt{H}}F(U,V)=0$. 

\section{Upper bound on optimal size for programmable detectors}

We will now derive an upper bound for the function $d=d(\varepsilon^{-1})$,
where $\epsilon=\min_{\bF}\varepsilon(\bF)$, that gives the minimal
needed dimension of the ancilla to achieve accuracy $\varepsilon^{-1}$
in programming observables for a finite-dimensional quantum system.
Clearly the function $d=d(\varepsilon^{-1})$ must be increasing, since
the higher is the accuracy $\varepsilon^{-1}$, the larger the minimal
dimension $d$ needed for the ancilla, namely the ``size'' of the
programmable detector.\par 

Consider now a $d$-dimensional ancilla and a system-ancilla
interaction $U$ of the following {\em controlled-unitary} form
\begin{equation}
U=\sum_{k=1}^dW_k\otimes|\phi_k\>\<\phi_k|,\label{cntrW}
\end{equation}
where $\{\phi_k\}$ is an orthonormal complete set of vectors for the
ancilla and $W_k$ are generic unitary operators on $\sH$. Consider
then a POVM $\bE=U\bF U^\dag$ of the form
\begin{equation}
E_i=|\psi_i\>\<\psi_i|\otimes I_A\,,
\end{equation}
where $I_A$ denotes the identity operator on the ancilla space, and
$\{\psi_k\}$ is a complete orthonormal set for the system. The
observable to be approximated will then be written as follows
\begin{equation}
P_i=W^\dag|\psi_i\>\<\psi_i|W,\label{obsW}
\end{equation}
$W$ being a unitary operator on $\sH$, and we will scan all possible
observables by varying $W$.  For the program state of the ancilla we
use one of the states $\phi_k$, which give the POVM's
\begin{equation}
Q_i=W_k^\dag|\psi_i\>\<\psi_i|W_k\,.
\label{eq:unitobs}
\end{equation}
This special form simplifies the calculation of the bound in Eq.
\eqref{eq:normbound}, which becomes
\begin{equation}
\begin{split}
\delta(\bP,\bQ)&\leqslant\sum_i\sqrt{2(1-|\<\psi_i|W^\dag W_k|\psi_i\>|^2)}\\
&\leqslant\sqrt 2\sum_i\sqrt{2-\<\psi_i|(W^\dag W_k-W^\dag_k W)|\psi_i\>},
\end{split}
\end{equation}
and using the Jensen's inequality for the square root function we have
\begin{equation}
\delta(\bP,\bQ)\leqslant\sqrt{2n}\N{W-W_k}_2\,.
\label{eq:finbound}
\end{equation}
Now we can always take $d$ sufficiently large such that we can choose
the $d$ operators $\{W_k\}$ in the unitary transformation $U$ in Eq.
(\ref{cntrW}) in such a way that for each given $W$ there is always a
unitary operator $W_k$ in the set for which $\sqrt{2n}\N{W-W_k}_2$ is
bounded by $\varepsilon$. This will guarantee that for the given
observable $\bP$ corresponding to $W$ there is a program state for the
ancilla such that the POVM $\bQ$ achieved by the programmable detector
is close to the desired $\bP$ less than $\varepsilon$. The set of all
possible unitary operators $W$ is a compact manifold of dimension
$h=n^2-n$. We now consider a covering of the manifold with balls of
radius $r=\frac{\varepsilon}{\sqrt{2n}}$ centered at the operators
$W_k$. This guarantees that any $W$ would be within a distance
$\frac{\varepsilon}{\sqrt{2n}}$ from an operator $W_k$, which in turns
implies that the accuracy of the programmable device is bounded by
$\varepsilon$ via Eq. \eqref{eq:finbound}. Using the volume
$V=\frac{\pi^{\frac{h}{2}}r^h}{\Gamma(\frac{1}{2}h+1)}$ of the
$h$-dimensional sphere of radius $r$, we obtain the number of balls
needed for the covering (for sufficiently small $\varepsilon$,
corresponding to the upper bound for the minimal dimension of the
ancilla
\begin{equation}
d\leqslant\kappa(n)\left(\frac1\varepsilon\right)^{n(n-1)}\,,
\label{eq:polyn}
\end{equation}
where $\kappa(n)$ is a constant that depends on $n$. Eq.
(\ref{eq:polyn}) gives an upper bound for the dimension $d$ which is
polynomial versus the accuracy $\varepsilon^{-1}$.\par 

For qubits, the observable has only two elements, $P_0=|\psi\>\<\psi|$
and $P_1=|\psi_\perp\>\<\psi_\perp|=I-P_0$, and the distance in Eq.
\eqref{eq:dist} can be evaluated analytically as follows
\begin{equation}
\delta(\bP,\bQ)=\max_\rho 2|\Tr[\rho(P_0-Q_0)]|\,.
\label{eq:distqub}
\end{equation}\par

As regards now the programmability of all POVM's (i.~e. including the
nonorthogonal ones), just notice that one just needs to be able to
program only the extremal POVM's in $\Pn$, since their convex
combinations will corresponds to mixing the program state or to
randomly choosing among different detectors. Then, since their maximum
number of outcomes is $n^2$, the extremal POVM's have Naimark's
extension to observables in dimension $n^2$, whence we are reduced to
the case of programmability of observables in dimension $n^2$.\par

We will now give a programmable detector for qubits that achieves an
accuracy that is linear in $d$. For the ancilla we use a generic
$d$-dimensional quantum system, and relabel the dimension in the
angular momentum fashion $d\doteq 2j+1$. The idea is now to design a
programmable detector in which the unitary transformation
corresponding to the observable $\{P_i\}$ in Eq. (\ref{obsW}) is
programmed by covariantly changing the program state of the ancilla.
By labeling unitary transformations by a group element $g\in\SU2$, we
write the observable to be programmed as $P_0\doteq
V_g|\frac12\>\<\frac12|V_g^\dag$ where $\{V_g\}\equiv(\tfrac{1}{2})$
is a unitary irreducible representation of $\SU2$ with angular
momentum $\tfrac{1}{2}$, whereas the program state will be written as
$W_g\sigma W_g^\dag$, with $\{W_g\}\equiv(j)$ a unitary irreducible
representation of $\SU2$ on the ancilla space with angular momentum
$j$. As already noticed, without loss of generality we can always
choose the state $\sigma$ as pure. We will now show that a good choice
for the program state is $\sigma=|j,j\>\<j,j|$, $\{|j,m\>\}$ denoting
an orthonormal basis of eigenstates of $J_z$ in the irreducible
representation with angular momentum $j$. The tensor representation
$\{V_g\otimes W_g\}\equiv\tfrac{1}{2}\otimes j$ can be decomposed into
the direct sum of two irreducible representations $\tfrac{1}{2}\otimes
j=j_+\oplus j_-$, where $j_\pm=j\pm \tfrac{1}{2}$. For the POVM $\bF$
of the programmable detector we will use $F_0=Z_+$ and $F_1=Z_-$,
$Z_\pm$ denoting the orthogonal projector on the invariant space for
angular momentum $j_\pm$
\begin{equation}
F_0=\sum_{m=-j_+}^{j_+}\left|j_+, m\right\>\left\<j_+, m\right|.
\end{equation}
Using the invariance $(V_g\otimes W_g)F_0(V_g^\dag\otimes
W_g^\dag)=F_0$, we can write the programmed POVM as follows
\begin{equation}
\begin{split}
Q_0=&\Tr_A[(I\otimes W^\dag_g|j,j\>\<j,j|W_g)F_0]
\\=&V_g^\dag\Tr_A[(I\otimes|j,j\>\<j,j|)F_0]V_g
\\ =&V_g\left(\left|\tfrac12,\tfrac12\right\>\left\<\tfrac12,\tfrac12\right|
+\tfrac1{2j+1}\left|\tfrac12,-\tfrac12\right\>\left\<\tfrac12,-\tfrac12\right|\right)V_g^\dag,
\end{split}
\end{equation}
where we used the only non vanishing Clebsch-Gordan coefficients
$|\<j_+,j_+|\tfrac12,\tfrac12\>|j,j\>|^2=1$, and
$|\<j_+,j_-|\tfrac12,-\tfrac12\>|j,j\>|^2=\frac{1}{2j+1}$. Clearly,
$Q_0-P_0=\frac1{2j+1}V_g|\frac12,-\frac12\>\<\frac12,-\frac12|V_g^\dag$,
whence according to Eq. (\ref{eq:distqub}) the accuracy is given by
$\delta(\bP,\bQ)=2/d$. The scaling of the dimension with the accuracy
is then linear
\begin{equation}
d=2\varepsilon^{-1},
\end{equation}
whereas the bound (\ref{eq:polyn}) would be quadratic
$d\propto\varepsilon^{-2}$. Sublinear growth of $d$ versus
$\varepsilon^{-1}$ is not excluded in general, but is not possible for
the present model.

\section*{Appendix: Derivation of the minimum fidelity for the unitary $V$}
\label{appbloch}

Starting from Eq.~\eqref{suvqub} we will obtain Eq.~\eqref{optV}. The
unitary $U$ belongs to $\SU2$, and can be written in the Bloch form
\begin{equation}\label{Bloch}
U=n_0I+i\vec n\cdot\vec\sigma\,,
\end{equation}
with $n_k\in\Reals$ and $n_0^2+|\vec n|^2=1$. Using the identity
\begin{equation}
\sigma_j \sigma_l \sigma_j=\epsilon_{jl}\sigma_l,\qquad
\epsilon_{j0}=\epsilon_{jj}=1,\quad\epsilon_{jl}=-1\,, l\neq 0,j,
\end{equation}
we can rewrite
\begin{equation}
S(U,V)=\tilde n_0 I+\tilde{\vec n}\cdot\vec\sigma,
\end{equation}
where 
\begin{equation}
\begin{split}
  \tilde n_j=&t_jn_j,\quad 0\leq j\leq 3,\quad t_0=\frac12\sum_{j=0}^3e^{-i\theta_j},\\
  t_j=&e^{-i\theta_0}+e^{-i\theta_j}-t_0,\;1\leq j\leq 3,\; 0\leq j\leq 3, \\
\end{split}
\end{equation}
and we will use the exponential representation for the complex number
$t_j=|t_j|e^{i\phi_j}$. It is now easy to evaluate the operator
$S(U,V)^\dag S(U,V)$. One has
\begin{equation}
\begin{split}
  &S(U,V)^\dag S(U,V)=v_0 I+ \vec v\cdot\vec \sigma,\\
  &v_0=|\tilde n_0|^2+|\tilde{\vec n}|^2,\quad \vec
  v=i\left[2\Im(\tilde n_0\tilde{\vec n}^*)+\tilde{\vec
      n}^*\times\tilde{\vec n}\right]\,.
\end{split}
\end{equation}
Now, the maximum eigenvalue of $S(U,V)^\dag S(U,V)$ is $v_0+|\vec v|$,
and one has
\begin{equation}
|\vec v|^2=2\sum_{i,j=0}^3|\tilde n_i|^2|\tilde n_j|^2\sin^2(\phi_i-\phi_j),
\end{equation}
whence the norm of $S(U,V)$ is given by
\begin{equation}\label{SUV}
\N{S(U,V)}^2=\vec u\cdot\vec t+\sqrt{ \vec u\cdot\vec T\vec u}\,,
\end{equation}
where $\vec u=(n_0^2 ,n_1^2 ,n_2^2 ,n_3^2 )$, $\vec t=(|t_0|^2
,|t_1|^2 ,|t_2|^2 ,|t_3|^2)$, and $\vec
T_{ij}=|t_i|^2|t_j|^2\sin^2(\phi_i-\phi_j)$. Notice that the unitary
$U$ which is programmed with minimum fidelity in general will not not
be unique, since the expression for the fidelity depends on
$\{n_j^2\}$. Notice also that using the decomposition in
Eq.~\eqref{parambipu} the minimum fidelity just depends on the phases
$\{\theta_j\}$, and the local unitaries will appear only in the
definitions of the optimal program state and of the worstly
approximated unitary. One has the following bound on the expression in
Eq.  (\ref{SUV})
\begin{equation}
\vec u\cdot\vec t+\sqrt{ \vec u\cdot\vec T\vec u}\geq \vec u\cdot\vec t\geq \min_j |t_j|^2, 
\end{equation}
and the bound is achieved on one of the four extremal points
$u_l=\delta_{lj}$ of the domain of $\vec u$ which is the convex set
$\{\vec u,\; u_j\geq 0,\,\sum_j u_j=1\}$ (the positive octant of the
unit four dimensional ball $S^4_+$). This proves the content of
Eq.~\eqref{optV}.\par

\bigskip
This work has been co-founded by the EC under the program ATESIT
(Contract No.  IST-2000-29681) and the MIUR cofinanziamento 2003
and FIRB 2004-2006. P.P.
acknowledges support from the INFM under project PRA-2002-CLON. G.M.D.
acknowledges partial support by the MURI program administered by the
U.S. Army Research Office under Grant No. DAAD19-00-1-0177.


\begin{thebibliography}{99}






\bibitem{Nielsen}M. A. Nielsen and I.  L. Chuang. \newblock {\em
    Quantum Computation and Quantum Information}. \newblock Cambridge
  University press, 2000.
\bibitem{dsw} G. M. D'Ariano, D. Kretschmann, D. Schlingeman, R. F. Werner, \newblock
  unpublished.
\bibitem{niels} M. A. Nielsen and I. L. Chuang. \newblock Programmable
  Quantum Gate Arrays. \newblock {\em Phys. Rev. Lett.}, 79:321--324,
  1997.
\bibitem{fiurasek2} J. Fiur\'asek and M. Du\v sek. \newblock
  Probabilistic quantum multimeters. \newblock {\em Phys. Rev. A},
  69:032302, 2004.
\bibitem{our} G. M. D'Ariano, P. Perinotti, and P. Lo Presti.
  \newblock Classical randomness in quantum measurements. \newblock
  {\em J. Phys. A: Math. Gen.}, 38:5979--5991, 2005
\bibitem{buzekproc} M. Hillery, V. Bu\v zek, and M. Ziman. \newblock
  Probabilistic implementation of universal quantum processors.
  \newblock {\em Phys. Rev.  A}, 65:022301, 2002.
\bibitem{buzek} M. Du\v sek and V. Bu\v zek. \newblock
  Quantum-controlled measurement device for quantum-state
  discrimination. \newblock {\em Phys. Rev. A}, 66:022112, 2002.
\bibitem{ciravid} G. Vidal and J. I. Cirac. \newblock Storage of
  quantum dynamics on quantum states: a quasi-perfect programmable
  quantum gate. \newblock quant-ph/0012067, 2000.
\bibitem{fiurasek1} J. Fiur\'a\v sek, M. Du\v sek, and R.  Filip.
  \newblock Universal Measurement Apparatus Controlled by Quantum
  Software. \newblock {\em Phys.  Rev. Lett.}, 89:190401, 2002.
\bibitem{raginski} M. Raginsky. \newblock A fidelity measure for
  quantum channels. \newblock {\em Phys. Lett. A}, 290:11--18, 2001.
\bibitem{KrausCirac} B. Kraus and J. I. Cirac. \newblock Optimal
  creation of entanglement using a two-qubit gate. \newblock {\em
    Phys. Rev. A}, 63:062309, 2001.
\bibitem{glaser} N Khaneja, R. Brockett, and S. Glaser. \newblock Time
  optimal control in spin systems. \newblock {\em Phys. Rev. A},
  63:032308, 2001.
\bibitem{last} G. M. D'Ariano and P. Perinotti. \newblock On the most
  efficient unitary transformation for programming quantum channels.
  \newblock quant-ph/0509183, 2005.
\end{thebibliography}
\end{document}